\documentclass[prd,aps,superscriptaddress,floatfix,sort&compress,nofootinbib,showpacs,twocolumn]{revtex4}
\usepackage{exscale}                  
\usepackage[intlimits]{amsmath}       
\usepackage{amsfonts}
\usepackage{amssymb,amscd}
\usepackage[dvips]{epsfig}
\usepackage{array}
\usepackage{wrapfig}
\usepackage{color}

\usepackage{amsfonts}
\usepackage{amsthm}
\usepackage{amssymb}
\usepackage{amsmath}
\usepackage{bm}
\usepackage{bbm}

\usepackage[colorlinks=true,linkcolor=blue,citecolor=blue,urlcolor=blue]{hyperref}

\newcommand{\beqa}{\begin{eqnarray}}
\newcommand{\eeqa}{\end{eqnarray}}
\newcommand{\beq}{\begin{equation}}
\newcommand{\eeq}{\end{equation}}

\newcommand{\qq}{\bar{q}q}

      \newcommand{\conjg}[1]{\ensuremath{\hspace{1pt}\overline{\hspace{-1pt}#1\hspace{-1pt}}}\hspace{1pt}}

         \newcommand{\be}{\begin{equation}}
         \newcommand{\ee}{\end{equation}}

\def\Slash#1{\setbox0=\hbox{$#1$} 
\dimen0=\wd0 
\setbox1=\hbox{/} \dimen1=\wd1 
\ifdim\dimen0>\dimen1 
\rlap{\hbox to \dimen0{\hfil/\hfil}} 
#1 
\else 
\rlap{\hbox to \dimen1{\hfil$#1$\hfil}} 
/ 
\fi}

\def\longlonglongrightarrow{
\relbar\joinrel\relbar\joinrel\relbar\joinrel\relbar\joinrel\relbar\joinrel\relbar\joinrel\rightarrow}

\begin{document}

\title{Tetraquark bound states in a Bethe-Salpeter approach}

\author{Walter Heupel}
\affiliation{Institut f\"{u}r Theoretische Physik,
Justus-Liebig-Universit\"at Giessen, D-35392 Giessen, Germany}

\author{Gernot Eichmann}
\affiliation{Institut f\"{u}r Theoretische Physik,
Justus-Liebig-Universit\"at Giessen, D-35392 Giessen, Germany}

\author{Christian~S.~Fischer}
\affiliation{Institut f\"{u}r Theoretische Physik,
Justus-Liebig-Universit\"at Giessen, D-35392 Giessen, Germany}
\affiliation{GSI Helmholtzzentrum f\"ur Schwerionenforschung GmbH,
  Planckstr. 1, D-64291 Darmstadt, Germany.}

\date{\today}

\begin{abstract}
We determine the mass of tetraquark bound states
from a coupled system of covariant Bethe-Salpeter equations.
Similar in spirit to the quark-diquark model of the nucleon, we
approximate the full four-body equation for the tetraquark
by a coupled set of two-body equations with meson and diquark
constituents. These are calculated from their quark
and gluon substructure using a phenomenologically well-established
quark-gluon interaction. For the lightest scalar tetraquark we find
a mass of the order of 400 MeV and a wave function dominated by the 
pion-pion constituents. Both results are in agreement with a meson 
molecule picture for the $f_0(600)$. Our results furthermore suggest 
the presence of a potentially narrow all-charm tetraquark in the 
mass region $5-6$ GeV.

\end{abstract}


\maketitle


    {\bf Introduction}\\
    The nature of the light scalar $0^{++}$ meson nonet 
    is still an issue under debate. Experimentally, the
    lowest-lying states in this channel are very broad which makes it
    difficult to identify their structure and properties. While
    other multiplets are firmly established to be $\qq$ mesons,
    the lightest scalar mesons $\sigma$, $\kappa$,
    $a_0(980)$ and $f_0(980)$ display signatures that signal a potential
    strong non-$\qq$ component. Once abandoned from the particle data book,
    these states were reintroduced only in the last decade
    due to new experiments such as KLOE in the
    $e^-e^+\rightarrow \pi^0\pi^0 \gamma$ channel~\cite{Caprini:2005zr}
    or BES in the $J/\Psi\rightarrow\omega\pi^-\pi^+$ channel~\cite{Ablikim:2004qna};
    see also~\cite{Nakamura:2010zzi} for a
    thorough compilation.

    Recently, several approaches
    to data analyses utilizing conventional and modified Roy equations
    have deduced a pole mass and width of the order of
    $m_\sigma\approx 450 + i 280$ MeV for the $\sigma/f_0(600)$
    \cite{Caprini:2005zr,GarciaMartin:2011jx}, with error bars below the
    five-percent level. Of course, these analysis do not answer the
    question on the nature of these states. In this respect, a number of
    arguments surfaced over the years which question the $\qq$ nature
    of light scalar mesons. In the non-relativistic quark model, scalar 
    quarkonia are $p$ waves, yet the $0^{++}$ nonet members are comparatively 
    light and the $f_0$(600) lies even below the $1^{--}$ isoscalar
    ($s-$wave) $\omega$ meson.
    The mass ordering within the $0^{++}$ nonet is puzzling as well: the 
    lowest-lying state is the isoscalar $f_0$(600) instead of the isotriplet 
    $a_0$(980), and the mass degeneracy of $a_0$(980) and $f_0$(980) is hard 
    to reconcile with their different flavor content. Finally, their decay 
    channels disagree with a $\qq$ picture: both $a_0$(980) and $f_0$(980) 
    couple to $K\bar{K}$ although only the latter contains strangeness, and 
    the broadness of the dominant decay channel $f_0(600)\rightarrow\pi\pi$ remains unexplained.

    The unexpected behavior of these light scalar mesons can be resolved
    in a tetraquark assignment which was introduced long ago by Jaffe \cite{Jaffe:1976ig}.
    Here, the oddities mentioned above are naturally explained by the flavor
    structure of the $qq\bar{q}\bar{q}$ nonet.
    The mass spectrum deduced from the tetraquark nonet is inverse to that
    of the $q\bar{q}$ nonet, thus explaining the low mass of the isoscalar.
    A $0^{++}$ tetraquark carries zero quark orbital angular momentum~\cite{Santopinto:2006my}, in
    agreement with the expectation that such $s$-wave states should be light.
    The putative decay channels of the tetraquark nonet agree with the
    observed ones, i.e., the coupling of $a_0$ and $f_0$ to $K\bar{K}$ and $\eta\pi$
    is caused by their strange-quark content, while the broadness of the $f_0(600)$
    is a consequence of its OZI-superallowed decay into two pions.
    These phenomenological arguments based on the group structure of the
    tetraquark flavor nonet are backed by effective theory studies and large-$N_c$
    arguments (see e.g. \cite{Achasov:1987ts,Black:1998wt,Maiani:2004uc,Giacosa:2006rg,Klempt:2007cp,Ebert:2008id} 
    and references therein) as well as recent
    lattice calculations~\cite{Alford:2003xw, Mathur:2006bs, Prelovsek:2010ty}
    which suggest a strong $qq\bar{q}\bar{q}$-component in the lowest-lying $0^{++}$ states.

    In this letter we present the first results for tetraquarks 
    in a covariant continuum approach based on the corresponding four-body
    equation for two quarks and two antiquarks \cite{Khvedelidze:1991qb}. We 
    construct a suitable representation of this system 
    in terms of mesons and diquark degrees of freedom. While we choose 
    this approximation for the sole reason of its numerical simplicity compared 
    to the full four-body equation, its feasibility is well motivated by recent 
    results in the baryon sector. There, the analogous quark-diquark approximation 
    to the nucleon's three-body Faddeev equation
    works well on the five-percent level~\cite{Eichmann:2008ef,Eichmann:2009qa,Eichmann:2011vu}.
    The relevance of diquark components in the systematics of hadron physics 
    has also been emphasized in Refs.~\cite{Jaffe:2005zz}. It is furthermore important
    to note, that the meson and diquark degrees of freedom that appear in our approach 
    are not fundamental; they are obtained dynamically from quark-(anti-)\-quark Bethe-Salpeter equations 
    (BSEs) and a model for the quark-gluon interaction which has been very successful 
    on a phenomenological level~\cite{Maris:2003vk}. We therefore determine the 
    properties of tetraquarks from the fundamental quark and gluon degrees of freedom of QCD.

    \begin{figure*}[t]
        \includegraphics[scale=0.088]{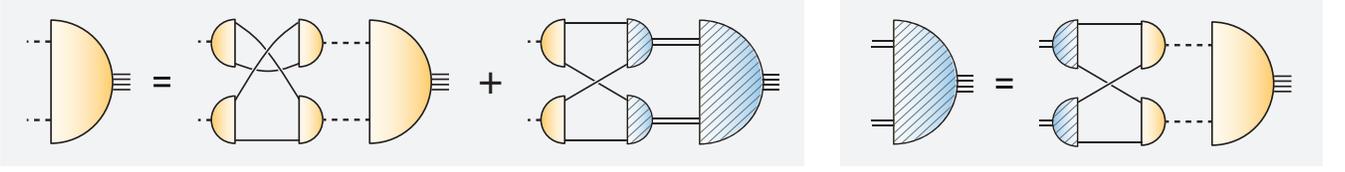}
        \caption{Tetraquark BSE in the meson-meson/antidiquark-diquark picture.
                 The hatched amplitudes involve diquark quantities; the remaining ones are of mesonic nature.
                 Single (double, dashed) lines are dressed quark (diquark, meson) propagators.}
        \label{fig:tetraquark}
    \end{figure*}


    {\bf The Bethe-Salpeter equation for tetraquarks}\\
    We start with the full four-quark Green function which satisfies the relation
    \begin{equation}
     G = G_0 +  G_0 \,T \, G_0.
      \label{eq:G_Matrix}
    \end{equation}
    Here, $G_0$ is the product of four dressed quark propagators and $T$
    represents the connected and fully amputated four-quark scattering matrix. In our
    symbolic notation the multiplications in Eq.~\eqref{eq:G_Matrix} represent
    four-dimensional integrations over the appropriate number of momenta and
    all indices are left implicit.
    Tetraquark bound states with mass $M$ appear
    as poles in the scattering matrix
    \begin{equation}
      T\stackrel{P^2=-M^2}{\longlonglongrightarrow} \frac{\Psi\,\conjg{\Psi}}{P^2+M^2}
      \label{poles-in-T}
    \end{equation}
    and thereby define the tetraquark's covariant bound-state amplitude $\Psi$,
    with $\conjg{\Psi}$ being its charge conjugate and $P$ the total momentum.

    The Green function satisfies a Dyson equation which relates it to the
    four-body interaction kernel $K$ via
    \begin{equation} \label{dyson-eq}
        G = G_0 + G_0\,K\,G \quad \Leftrightarrow \quad T = K + K\,G_0\,T\,.
    \end{equation}
    Substituting Eq.~(\ref{poles-in-T}) into \eqref{dyson-eq} and projecting
    onto the singular part, one arrives at a homogeneous bound-state equation for the tetraquark amplitude:
    \begin{equation}
     \Psi=K\,G_0\,\Psi.
    \label{eq:BSE_4b}
    \end{equation}
    This equation can be solved once the four-body kernel $K$ is known.
    The kernel contains 2PI, 3PI and 4PI contributions~\cite{Khvedelidze:1991qb}; in the following we adopt the successful
    strategy of Ref.~\cite{Eichmann:2009qa} and neglect the latter two.

    To simplify the notation, we suppress the quark propagators
    by replacing $G_0^{-1} G \rightarrow G$, $T \,G_0 \rightarrow T$ and $K \,G_0 \rightarrow K$,
    so that Eqs.~\eqref{eq:G_Matrix}, \eqref{dyson-eq} and \eqref{eq:BSE_4b} become
    \begin{equation*}
        G = 1 + T = 1 + K\,G\,, \quad  T = K\,(1+T)\,,  \quad \Psi = K \,\Psi\,.
    \end{equation*}
    The remaining part of the kernel that involves only two-body correlations can be written as the sum of three terms:
    \begin{equation}\label{2-body-kernel}
       K = \sum_{aa'} K_{aa'}\,,
    \end{equation}
    where $a$ and $a'$ denote $qq$, $\bar{q}\bar{q}$ or $q\bar{q}$ pairs,
    so that $aa'$ is one of the three combinations $(12)(34)$, $(13)(24)$ or $(14)(23)$.
    Therefore, $K_{aa'}$ describes the component of the four-body kernel where all interactions are switched off
    except those within the pairs $a$ and $a'$.

    The absence of residual color forces between widely separated clusters,
    potentially generated by the permuted interactions in Eq.~\eqref{2-body-kernel},
    amounts to the condition
    \begin{equation}\label{G-separability}
       G_{aa'} = G_a\,G_{a'}\,.
    \end{equation}
    Here, $G_{aa'}=1 + T_{aa'}$ is the four-body Green function obtained from the kernel $K_{aa'}$,
    whose scattering matrix satisfies
    \begin{equation}\label{T1}
       T_{aa'} = K_{aa'}(1+T_{aa'})\,,
    \end{equation}
    and $G_a=1+T_a$, $G_{a'}=1+T_{a'}$ are the two-body Green functions for the individual pairs $a$ and $a'$, with
    \begin{equation}\label{T2}
       T_a = K_a\,(1+T_a)\,, \quad T_{a'} = K_{a'}\,(1+T_{a'})\,.
    \end{equation}
    The separability of the Green function $G_{aa'}$ imposes the following structure on the kernel $K_{aa'}$~\cite{Khvedelidze:1991qb}:
    \begin{equation}\label{2-body-kernel-1}
       K_{aa'} = K_a + K_{a'} - K_a\,K_{a'}\,,
    \end{equation}
    where $K_a$ and $K_{a'}$ are now elementary $qq$, $\bar{q}\bar{q}$ or $q\bar{q}$ kernels.
    The relations (\ref{T1}--\ref{2-body-kernel-1}) yield the scattering matrix
    \begin{equation}\label{T-matrix}
       T_{aa'} = T_a + T_{a'} + T_a\,T_{a'}\,,
    \end{equation}
    from which Eq.~\eqref{G-separability} can be readily verified.

    In principle, the tetraquark bound-state equation \eqref{eq:BSE_4b} with the kernel of
    Eqs.~\eqref{2-body-kernel} and~\eqref{2-body-kernel-1} can be solved with the techniques used in
    Ref.~\cite{Eichmann:2009qa} for the three-body equation. In practice,
    however, this is a very demanding task in terms of computation power.
    While we strive to attack this problem in the future, for now we resort
    to a further, simplifying approximation in the spirit of the nucleon's Faddeev equation and its reduction to a quark-diquark picture~\cite{Eichmann:2008ef}.
    In analogy to the three-body case, we define Faddeev amplitudes $\Psi_{aa'}$ via
    \begin{equation}
        K_{aa'} \,\Psi =: \Psi_{aa'} \quad \Rightarrow \quad \Psi = \sum_{aa'} \Psi_{aa'}\,.
    \end{equation}
    Upon projecting Eq.~\eqref{T1} onto the bound-state amplitude $\Psi$, one obtains Faddeev-Yakubovsky type equations~\cite{Yakubovsky:1966ue}
    for the amplitudes $\Psi_{aa'}$:
    \begin{equation}\label{faddeev}
        \Psi_{aa'} = T_{aa'}\,(\Psi_{bb'} + \Psi_{cc'})\,, \quad aa' \neq bb' \neq cc'\,,
    \end{equation}
    where $T_{aa'}$ is constructed from two-body scattering matrices according to Eq.~\eqref{T-matrix}.

    Except for the omission of genuine three- and four-body correlations, Eq.~\eqref{faddeev} is still exact.
    Its reduction to a two-body problem proceeds by
    assuming that the two-body $T-$matrices are dominated by meson and diquark pole contributions,
    \begin{equation}
     T_a(q_1,q_2,Q)=-\Gamma_a(q_1,Q) \,D_a(Q) \,\Bar{\Gamma}_a(q_2,Q)\,,
    \label{eq:T_matrix_pole_assumption}
    \end{equation}
    and that the internal spin-momentum structure of the Faddeev amplitudes factorizes:
    \begin{equation}
    \begin{split}
     \Psi_{aa'}(p,q,q',P) &= \Gamma_a(q,Q)\,D_a(Q)   \\
                             & \times \conjg{\Gamma}_{a'}(q',Q') \,D_{a'}(Q')\,\Phi_{aa'}(p,P).
    \label{eq:Amplitude_Sep_Tetra}
    \end{split}
    \end{equation}
    Without loss of generality we can assign the labels $1, 2$ to the quarks and $3, 4$ to the antiquarks.
    Then, for $aa'=(12)(34)$, $\Gamma_a$ and $\conjg{\Gamma}_{a'}$ describe
    diquark and antidiquark bound-state amplitudes and $D_a$, $D_{a'}$ their respective propagators, whereas
    in the case of $aa'= (13)(24)$ or $(14)(23)$ the involved objects are
    of mesonic nature. Quantities with a bar indicate charge-conjugated amplitudes.
    $P$ is the total tetraquark momentum and $p$ the relative momentum between its respective constituents.
    The separated internal momenta $q$, $q'$
    correspond to the relative momenta of the (anti-)diquarks and mesons.
    Their total momenta $Q$, $Q'$ can take arbitrary values and thus
    an off-shell description for the meson and diquark amplitudes is necessary.

    Combining Eqs.~(\ref{faddeev}--\ref{eq:Amplitude_Sep_Tetra}) and furthermore neglecting
    the single-interaction contributions $T_a$ and $T_{a'}$ from Eq.~\eqref{T-matrix}
    yields a coupled diquark-antidiquark/meson-meson
    BSE which is depicted in Fig.~\ref{fig:tetraquark}. It describes an effective interaction between
    two mesons, or between a diquark and an antidiquark, via quark exchange.
    We take into account the mesons and diquarks
    with lowest mass, i.e., the pseudoscalar-meson and scalar-diquark channels.
    Constituents with other quantum numbers are certainly possible; however, since the
    corresponding calculations are very expensive in terms of CPU
    time we postpone their inclusion to subsequent work.
    Moreover, we only investigate the lowest-lying tetraquark with quantum
    number $J^P=0^+$.
    The resulting diquark-antidiquark and meson-meson contributions to the tetraquark amplitude,
    \begin{equation}
    \begin{split}
        \Phi_\text{D}(p,P) &:= \Phi_{(12)(34)}\,, \\
        \Phi_\text{M}(p,P) &:= \Phi_{(13)(24)} = -\Phi_{(14)(23)} \,,
    \end{split}
    \end{equation}
    are flavor and color singlets and Lorentz scalars.

    It is noteworthy that our framework, Fig.~\ref{fig:tetraquark}, does not 
    permit a pure diquark-antidiquark state in isolation; it can only occur 
    in combination with meson-meson interactions. On the other hand, both 
    equations can be merged to a single mesonic equation where diquarks 
    appear only internally. Thus, one may view the resulting tetraquark bound 
    state as a meson molecule with diquark-antidiquark admixture to its kernel.
    We expect this diquark admixture to be especially important for tetraquarks
    with masses larger than the sum of their meson constituents.  


    {\bf Mesons and diquarks from quark and glue}\\
    In order to solve the tetraquark BSE of Fig.~\ref{fig:tetraquark}, we need to determine the
    (on-shell) masses and amplitudes of the meson and diquark building
    blocks as well as a suitable continuation off their mass shells.
    For diquarks, this problem has been dealt with already within the
    quark-diquark approach to the baryon three-body problem
    \cite{Eichmann:2008ef}. The corresponding techniques are well developed
    and their reliability can be judged from the good agreement of
    nucleon and $\Delta$ masses in the quark-diquark picture with results
    from the corresponding three-body problem~\cite{Eichmann:2009qa,SanchisAlepuz:2011jn}.
    We therefore adopt this framework also for the diquark and meson
    amplitudes that appear in our setup. The technical details of these
    types of calculations have been described in many works, see e.g.
    \cite{Maris:2003vk,Fischer:2006ub,Eichmann:2009zx} for reviews,
    thus we only give a short summary here.

    The starting point is the Dyson-Schwinger equation for the dressed quark propagator,
      \begin{equation}\label{quarkdse}
         S^{-1}_{\alpha\beta}(p) = Z_2 \left( i\Slash{p} + m_0 \right)_{\alpha\beta}
         + \int_q \mathcal{K}_{\alpha\alpha'\beta'\beta} \,S_{\alpha'\beta'}(q)\,,
      \end{equation}
    with wave-function renormalization constant $Z_2$ and bare quark mass $m_0$.
    The exact interaction kernel $\mathcal{K}_{\alpha\alpha'\beta'\beta}$ contains the dressed gluon propagator
    as well as one bare and one dressed quark-gluon vertex.
    The Greek subscripts refer to color, flavor and Dirac structure.
    In the rainbow-ladder approximation that we adopt here the kernel can be written as
        \begin{equation}\label{RLkernel}
            \mathcal{K}_{\alpha\alpha'\beta\beta'} =  Z_2^2 \, \frac{ 4\pi \alpha(k^2)}{k^2} \,
                 T^{\mu\nu}_k \gamma^\mu_{\alpha\alpha'} \,\gamma^\nu_{\beta\beta'},
        \end{equation}
    where $T^{\mu\nu}_k=\delta^{\mu\nu} - k^\mu k^\nu/k^2$  is a transverse projector
    with respect to the gluon momentum $k$. Eq.~\eqref{RLkernel} describes an iterated
    dressed-gluon exchange between quark and antiquark that retains only the vector
    part $\sim \gamma^\mu$ of the quark-gluon vertex. Its non-perturbative dressing,
    together with the one for the gluon propagator, is absorbed into an effective
    coupling $\alpha(k^2)$ which is taken from Ref.~\cite{Maris:1999nt,Eichmann:2009qa}.

    Chiral symmetry and the associated axial-vector Ward-Takahashi identity 
    demand the kernel $\mathcal{K}$ to appear in the corresponding
    BSEs as well. The meson BSE is given by
    \begin{align}
       \Gamma_{\alpha\beta}(p,P) =& \int_q K_{\alpha\alpha'\beta'\beta}
        \left\{S(q_+) \,\Gamma(q,P) S(q_-)\right\}_{\alpha'\beta'} ,
    \end{align}
    with the Bethe-Salpeter amplitude $\Gamma(p,P)$ depending on
    total and relative momenta of the quark and anti-quark constituents, and $q_\pm = q \pm P/2$.
    The corresponding equation for diquarks is obtained by the substitution
    of an antiquark with a quark leg. Both meson and diquark amplitudes
    contain four different Dirac structures which we compute explicitly.

\begin{figure*}[t]
\includegraphics[scale=0.34]{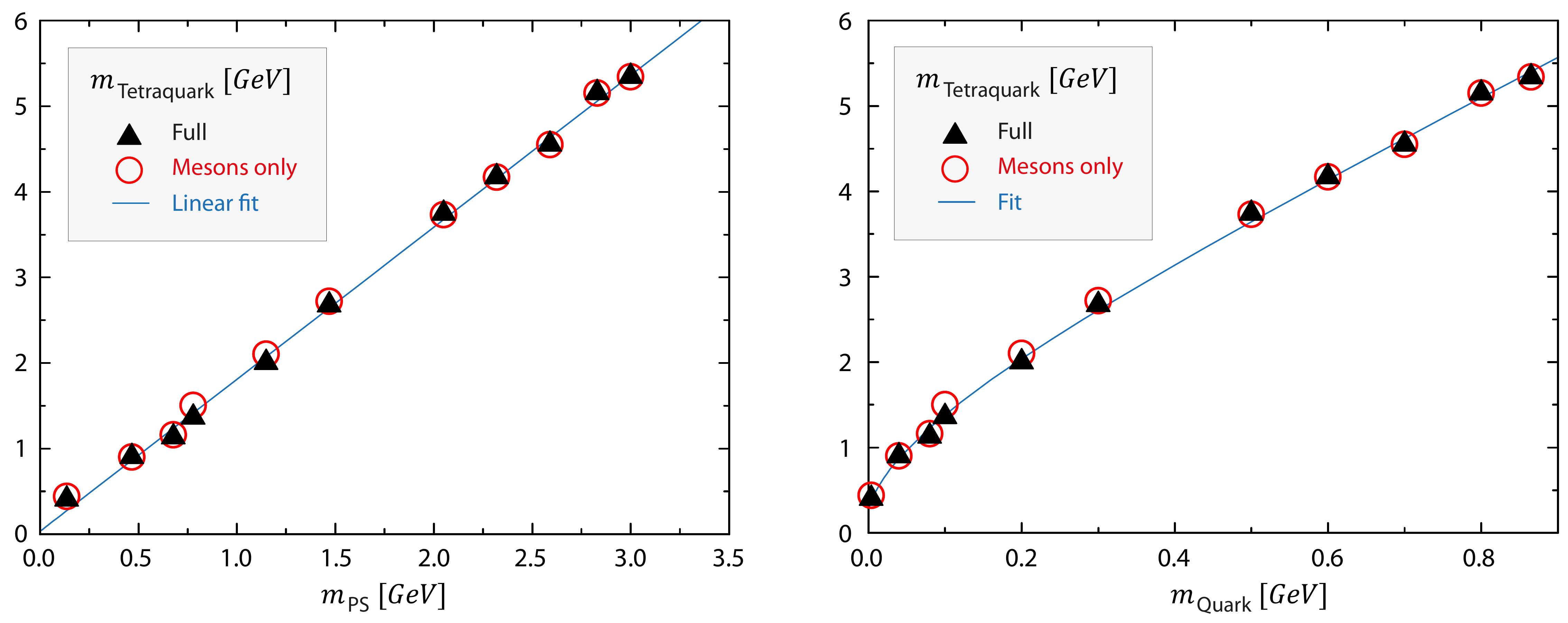}
\caption{Mass of the $0^{++}$ tetraquark as a function of the
pseudoscalar-meson mass (\textit{left}) and the quark mass (\textit{right panel}).}
\label{res:fig2}
\vspace*{0mm}
\end{figure*}

    With a given effective coupling, one determines the quark propagator in the
    complex momentum plane and subsequently solves the meson and diquark BSEs.
    According to the techniques developed in
    Refs.~\cite{Eichmann:2008ef,Eichmann:2009zx},
    the resulting onshell wave functions are analytically
    continued to offshell momenta and the effective meson and diquark
    propagators are computed from their $T-$matrix relations.
    Finally, all building blocks
    are put together in the tetraquark BSE.
    Our numerical techniques used to
    solve this BSE are only slightly non-standard and will be described in
    detail elsewhere.


{\bf Results and discussion}\\
Our result for the mass of the up/down $0^{++}$ tetraquark state as a function
of the pseudoscalar-meson mass is shown in the left panel of Fig.~\ref{res:fig2}, together with
a calculation that only includes the meson-molecule component of the tetraquark.
Clearly, the meson-meson component dominates.
Except for the chiral region, the overall dependence of the tetraquark mass upon the
pseudoscalar-meson mass is linear within numerical errors, as
can be seen from the comparison with the linear fit included in the plot. The reason
for this behavior is even more clear from the right panel of Fig.~\ref{res:fig2}, where we plot the
tetraquark mass as a function of the quark mass. The line represents
a fit to the data including a constant, a square root and a linear term.
Apart from the constant term this is the typical behavior of a Goldstone boson.
The tetraquark thus grossly inherits the mass behavior of its dominating pion-molecule
constituents, with deviations generated from their interactions via quark
exchange.

One of the main results of our present work is the value
for the u/d tetraquark at the physical point, i.e. the left-most
points in Fig.~\ref{res:fig2}, where $m_{\mbox{\tiny PS}} = m_\pi$.
We obtain:
\beq
m_\text{Tetraquark}^{u/d}(0^{++}) = 403 \,\mbox{MeV}\,,
\eeq
with an estimated numerical error of ten percent.
This value is only somewhat lower than the real part of the mass of the
$\sigma/f_0(600)$, $m_\sigma\approx 450 + i 280$ MeV determined
recently from experiment using Roy equations \cite{Caprini:2005zr,GarciaMartin:2011jx}. Our value
for the mass of the scalar tetraquark should also be compared with the
corresponding one for an ordinary quark-antiquark scalar bound state
which may mix with the tetraquark components.
In our rainbow-ladder approximation such a state has a mass of
$m_{q\bar{q}}(0^{++}) = 665 \,\mbox{MeV}$. It is well known that
corrections beyond rainbow-ladder increase this value into the
1 GeV range \cite{Fischer:2009jm,Chang:2011ei}, whereas
the pion mass is protected. Since our tetraquark
is dominated by its meson-molecule nature, we therefore expect it
to be stable against corrections beyond rainbow-ladder,
while at the same time the mass splitting between the tetraquark and
the quark-antiquark scalar will increase. Consequently, our results
suggest to identify the physical lowest-lying scalar state
to be dominated by a strong tetraquark component, which is in turn
dominated by pion molecule contributions. Due to the (Pseudo-) Goldstone 
nature of the pion constituents, our result provides a ready and 
natural explanation for the small mass and the large decay width 
of the $\sigma/f_0(600)$. 

In the strange quark region at about $m_\text{Quark}=80$ MeV we also observe 
an all-strange tetraquark bound state at roughly 
$m_\text{Tetraquark}^{s}(0^{++}) = 1.2 \,\mbox{GeV}$. Certainly this state
will mix with its pure $s\bar{s}$ counterpart as well as the lowest lying 
scalar glueball making an identification with $f_0(1500)$ or 
$f_0(1710)$ not possible without further studies. 

It is furthermore interesting to speculate about the
existence of an all-charm tetraquark state.
Because of its
flavor-structure in our meson-diquark picture, such a state would
be a mixture between a meson and an axialvector-diquark component.
Since already the scalar-diquark contribution is very small, we expect
the axialvector component to be completely suppressed due to its larger mass.
This leaves only the dominant meson-molecule part. In Fig.~\ref{res:fig2}
the largest pseudoscalar-meson mass corresponds to a quark mass in
the charm region. We therefore read off the mass of an all-charm scalar
tetraquark state to be at
\beq
m_\text{Tetraquark}^{c}(0^{++}) = 5.3 \pm (0.5) \,\mbox{GeV}\,,
\eeq
where the error is a guess based on our numerical and systematic uncertainties. This
mass is considerably lower than the $6.2$ GeV obtained in simple model
calculations \cite{Iwasaki:1975pv,Lloyd:2003yc}. It is also much lower
than the $\eta_c$ threshold. Potential decay channels into $D$ mesons
and pairs of light mesons necessarily involve internal gluon lines.
The resulting decay width may therefore be rather small.

Further results for tetraquark states with unequal mass constituents are 
numerically more demanding than the ones presented here and will only be
available for a future publication.

{\bf Acknowledgments}\\
We are grateful to Francesco Giacosa, Andreas Krassnigg, Soeren Lange and Milan Wagner for discussions.
This work was supported by the Austrian Science Fund FWF under
Erwin-Schr\"odinger-Stipendium No.~J3039, the Helmholtz International
Center for FAIR within the LOEWE program of the State of Hesse,
the Helmholtz Young Investigator Group under contract VH-NG-332 and BMBF under
contract 06GI7121.

\end{document}